\def\aa{A\&A}
\def\asr{Adv. Space Res.} 
\def\apj{ApJ} 
\def\apjl{ApJL} 
 
\def\apjs{ApJS}
\def\mnr{MNRAS}
 
\def\nat{Nat}

\def\mchisq{$\chi^2$} 
\def\mchisqr{$\chi^{2}_{\nu}$}

\def\mnu{$ \nu $}
\def\x{\times} 
\def\G{\Gamma}

\def\mnh{$N_{\rm H}$}
\def\nh{N_{\rm H}}
\def\funits{\rm ~erg~cm^{-2}s^{-1}} 
\def\lunits{\rm ~erg~s^{-1}} 
\def\nunits{\rm ~cm^{-2}} 

\newcommand{\simlt}{{\small\raisebox{-0.6ex}{$\,\stackrel
{\raisebox{-.2ex}{$\textstyle <$}}{\sim}\,$}}}
\newcommand{\simgt}{{\small\raisebox{-0.6ex}{$\,\stackrel
{\raisebox{-.2ex}{$\textstyle >$}}{\sim}\,$}}}   
   
\def\references{\subsection*{REFERENCES} 
\bgroup\parindent=0pt\parskip=\itemsep 
\def\refpar{\par\hangindent=1.2em\hangafter=1}} 
\def\endreferences{\refpar\egroup} 
\def\refpar{\relax} 
\def\re{\relax\refpar} 
\documentstyle[graphicx]{mn}

\begin{document} 

\title[The average QSO spectrum and its evolution]{A deep {\it ROSAT} survey
XV. The average QSO spectrum and its evolution.}
\author[A. J. Blair et al.]
{\large A. J. Blair,$^1$ G. C. Stewart,$^1$ I. Georgantopoulos,$^2$ B. J. Boyle,$^3$
 R. E. Griffiths,$^4$ T. Shanks,$^5$ O. Almaini\ $^6$\\
$^1$ Department of Physics and Astronomy, University of Leicester,   
     Leicester, LE1 7RH \\
$^2$ Astronomical Institute, National Observatory of Athens, Lofos
     Koufou, Palaia Penteli, Athens, GR-15236, Greece \\
$^3$ Anglo-Australian Observatory, PO Box 296, Epping NSW 2121, Australia \\
$^4$ Department of Physics, Carnegie Mellon University, Wean Hall, 
     5000 Forbes Ave., Pittsburgh, PA 15213 U.S.A. \\ 
$^5$ Physics Department, University of Durham, South Road, 
     Durham DH1 3LE \\
$^6$ Institute for Astronomy, University of Edinburgh, Blackford Hill, 
Edinburgh, EH9 3HJ}
\maketitle

\begin{abstract}

Using a sample of 165 X-ray selected QSOs from seven deep {\it ROSAT}
fields ($ f_{(0.5-2.0~{\rm keV})} \simgt 4 \x 10^{-15} \funits $), we
investigate the X-ray spectral properties of an ``average''
radio-quiet broad-line QSO as a
function of redshift. The QSO stacked spectra, in the observers $
0.1-2 $ keV band, in five redshift bins over the range $0.1 \le z \le
3.2 $ apparently  harden from an equivalent photon index of 
$\G \sim 2.6$ at $z=0.4$
to $\G \sim 2.1$ at $z=2.4$ as  seen in
other QSO samples. In contrast, the  spectra 
 in the 0.5-2 keV band show  no significant variation in spectral index 
 with redshift. This suggests the presence of a spectral upturn 
 at low energies ($<0.5$ keV). Indeed,   
 while at high redshifts ($z>1.0$) the single
 power-law model gives an acceptable fit
 to  the data over the full energy band, at lower
 redshifts the spectra need a second component at low energies, a
`soft excess'. 
 Inclusion of a
 simple model for the soft excess, i.e. a 
black-body component (kT$\sim 100$ eV), 
results in a significant improvement to the model fit, and yields
power-law slopes of $\G \sim 1.8-1.9$, for all redshift bins.  This
power-law is not inconsistent, within the error bars, with those of
nearby AGN in the $2-10$ keV band, suggesting that the same intrinsic
power-law slope may continue from 10 keV down to below $\sim$ 0.5 keV.
 We caution that there is a possibility that the 
 spectral upturn observed may not represent a real physical component 
 but could be  due to  
 co-adding spectra with a large dispersion in spectral indices. 
 Regardless of the origin of the soft excess, the average 
 QSO spectrum  has 
 important consequences for the origin of the X-ray background: 
 the average spectra of the {\it typical}, faint, high redshift QSO are 
 significantly steeper than the spectrum of the X-ray background 
 extending the spectral paradox  into the soft 0.1-2 keV X-ray band. 

\end{abstract}

\begin{keywords}
Active Galaxies: X-ray Spectra - Active Galaxies: Evolution
\end{keywords}

\section{Introduction}
Following the discovery of X-ray luminous QSO 
 $\sim25$ years ago (Lampton et al. 1972), we remain without a
detailed description of the form and emission mechanism of QSO X-ray
spectra, despite extensive studies of these objects.

Satellite missions such as  {\it Ginga} and {\it EXOSAT} indicated that
the X-ray emission of radio-quiet QSO
 (which constitute the vast majority of the QSO population)  
 above $\sim 2$ keV can be well described by
a power-law with a photon index of $\G \sim 1.9$ 
 (Williams et al. 1992, Lawson et al. 1992). However, the
sources detected by these experiments which are the X-ray brightest,
typically with $f_{\rm x}  \ge 1 \x 10^{-12}
\funits$, nearby ($z<0.2$)  QSO may not be representative of the 
class as a whole. Reeves et al. (1997) 
 observed a small number of bright radio-quiet QSOs 
 with {\it ASCA}, extending these results to higher redshifts.   
 Recent long-exposure ($\sim 100$ ksec) {\it ASCA}
observations (Georgantopoulos et al. 1997) extend the investigation of
the average QSO X-ray spectra down to even fainter flux limits of $\sim 5 \x
10^{-14} \funits$ in the $2-10$ keV band. Although, 
 the photon statistics are not sufficient to be conclusive,
 there is some evidence that the spectral index of these 
 faint QSOs  is 
 flat ($\G = 1.5 \pm 0.2$).

 Observations of the soft X-ray spectrum ($<2$ keV) of
radio quiet, optically selected QSOs made with  the {\it Einstein}
observatory show a  steeper spectral index ($\G > 2$).
 This suggests that the QSO spectra are concave i.e. they 
 cannot be represented by 
 a single power-law over a broad energy band (eg Schwartz \& Tucker 1988). 
 This steep soft spectral index could be 
 interpreted  as excess  emission (soft excess), at energies below $\sim 1$
keV, above the extrapolation of the X-ray power-law slope found at
higher energies. Such a soft excess is 
 detected  in a substantial number of quasars (Masnou et
al. 1992, Saxton et al. 1993), and in about 50 per cent of nearby
Seyfert type AGN (Turner \& Pounds 1989).  
Ciliegi \& Macaccaro (1996) examine the X-ray spectral
properties of 63 {\it X-ray selected} AGN, from the {\it Einstein} Medium
Sensitivity Survey, finding $\G \sim 2.4$ with an intrinsic dispertion
in spectral slope of $\sigma \sim 0.4$, suggesting that X-ray spectral
slopes are independent of whether selection takes place at either
X-ray or optical wavelengths. After the launch of {\it
ROSAT} (Tr\"umper 1982), sensitive in the range $0.1-2.4$ keV and
carrying instruments with improved energy resolution over previous
missions, information on the soft X-ray properties of QSO burgeoned,
due, in part, to large surveys with flux limits ($\sim 5 \x 10^{-15}
\funits$) orders of magnitude fainter than those of previous surveys. 
 Schartel et al. (1996a) studied the stacked (average) 
 spectra of a sample of 908 objects from the 
LBQS sample of optically selected QSOs. They find a mean
spectral slope of $\G \sim 2.5$ and marginal evidence for a flattening
of the spectral index at higher redshifts. Schartel et al. 
 (1996b) analysed the individual spectra of a sample of 55 radio-quiet 
 QSOs from the {\it ROSAT} all-sky survey. They find again 
 steep spectral indices $\Gamma \approx 2.5$
 and no  evidence for spectral evolution with redshift. 
 Laor al. (1997) also derive a steep spectra index $ \G \approx 2.72 \pm
0.09$ using a sample of 19 radio-quiet, optically selected QSOs with
$z<0.4$.

The above values for the spectral slope of QSOs constitute the
discrepancy known as the spectral paradox (Boldt 1987), in that the
most common class of X-ray source has a spectral index significantly
steeper than that of the X-ray background (XRB), which has $\G \sim
1.5$ at both hard  (Marshall et al, 1980, Gendreau et al. 1995)
and soft energies
(Georgantopoulos  et al. 1996), and so cannot be the major contributor to
the XRB flux. 
A resolution of the spectral paradox could be the
evolution of the mean QSO X-ray spectrum with redshift. 

Here, we  derive the average QSO spectral index  
 using a sample of X-ray selected QSOs detected in 
 seven fields from our deep {\it ROSAT} survey (Shanks et al. 1991,
 Georgantopoulos et al. 
 1996). Previous work by Almaini et al. (1996) in five of our fields 
 finds no variation with
redshift in the slope of the spectra of QSOs (as determined by
analysis of hardness ratios in the $0.5-2.0$ keV band) from five deep
{\it ROSAT} fields.  In order to identify the effect of a soft excess
(emission significant only at $\simlt0.5$ keV) on the resultant
power-law slope in the $0.1-2.0$ keV band, we  use here the full {\it
ROSAT} energy band and also perform proper spectral fits to the data.
In this paper we include two further  deep {\it ROSAT} fields in our
investigation, creating a sample of 165 QSOs covering a 
 broad redshift range, 
$0.1 \le z \le 3.2$, with a relatively high mean redshift ($\bar{z}
\approx 1.5$). We also include two longer expsores (50 ksec
each) of the QSF1 and QSF3 fields, which, along with the increased
range of energies considered, results in photon statistics that are
effectively doubled over those in the Almaini et al. work.
Preliminary results from this work were presented in Stewart et
al. (1994). The goals are to a) detect and quantify the possible evolution of 
 the QSO spectral index b) compare the QSO spectra with that of the 
 XRB c) examine the possibility for an upturn  of  the average spectral index 
 in the soft band. 

\section{X-ray data selection}

\begin{figure}
\rotatebox{270}{\includegraphics[height=7.5cm]{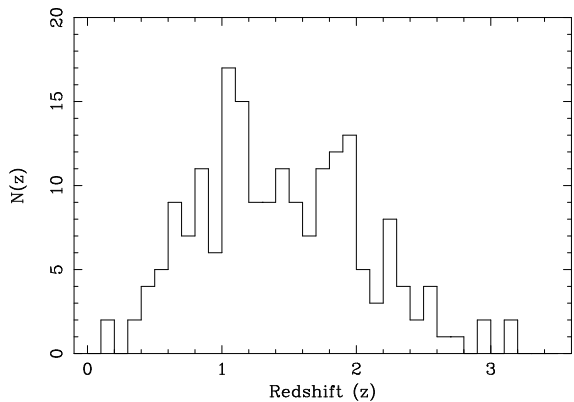}}
\caption{Redshift distribution, N(z), for the QSO sample, with a mean
redshift $\bar{z} = 1.44$.}
\label{fig:nz}
\end{figure}

We use data taken from deep observations of seven fields with the {\it
ROSAT} PSPC (a summary of the observations taken is shown in
table~\ref{tab:fields}). Over 300 sources were detected within the
inner-ring of the detector, at the $4 \sigma$ level, down to a flux of
$\sim 4 \x 10 ^{-15} \funits (0.5-2.0$ keV). After an optical
identification program with the AAT, 165 QSOs are identified, 
 see Georgantopoulos et al. (1996) for the full 
 details of the identification procedure and 
Shanks et al. (in preparation) for the catalogue of sources.
 Redshifts are in the range $0.1 \le z
\le 3.2$. Fig.~\ref{fig:nz} shows the distribution.

\begin{table}
\begin{center}
\begin{tabular}{cllcc}
Field  & \multicolumn{2}{c}{Pointing direction}& \mnh &Total \\
       & \multicolumn{2}{c}{(J2000)}& ($10^{20} \nunits$)&Exposure \\ \hline
BJS855 &10 46 24.0 &  $-00$ 01 58.8 & 3.8 & 11941 \\
BJS855 &10 46 24.0 &  $-00$ 01 58.8 & 3.8 & 16421 \\
BJS864 &13 43 43.2 &  $+00$ 15 00.0 & 2.8 & 23213 \\
GSGP4  &00 57 28.8 &  $-27$ 38 24.0 & 1.9 & 47426 \\
SGP2   &00 52 04.8 &  $-29$ 05 24.0 & 1.9 & 23758 \\
SGP3   &00 55 00.0 &  $-28$ 19 48.0 & 1.9 & 21837 \\
QSF1$^{\dagger}$   &03 42 09.6 &  $-44$ 54 36.0 & 1.6 & 26082 \\
QSF1   &03 42 09.6 &  $-44$ 54 36.0 & 1.6 & 48629 \\
QSF3$^{\dagger}$   &03 42 14.4 &  $-44$ 07 48.0 & 1.7 & 27270 \\
QSF3   &03 42 12.0 &  $-44$ 07 48.0 & 1.7 & 51918 \\
\end{tabular}
\caption{Details of observations used. The observations denoted by
$^{\dagger}$ were made with the PSPC-C instrument, all others were
made with the PSPC-B.}
\label{tab:fields}
\end{center}
\end{table}

\begin{table}
\begin{center}
\begin{tabular}{cccrc}
  &   Redshift   &$\bar{z}$&QSOs&Photons\\
  &   range      &         &    & $0.1-2.0$ keV\\ \hline
1 &  $0.0 - 0.5$ & $0.36$  & 7  & 6783\\	   
2 &  $0.5 - 1.0$ & $0.78$  & 32 & 4735\\	   
3 &  $1.0 - 1.5$ & $1.23$  & 54 & 5084\\	   
4 &  $1.5 - 2.0$ & $1.77$  & 44 & 3530\\	   
5 &  $2.0 - 3.2$ & $2.42$  & 28 & 1518\\	   
\end{tabular}
\caption{Details of redshift bins.}
\label{tab:bins}
\end{center}
\end{table}

Datasets used in this work are available from the LEDAS {\it ROSAT} public
database.  We exclude data from periods of high particle background
i.e. when the Master Veto Rate is above $170\ \rm count~s^{-1}$
(Plucinsky et al. 1993), typically $\sim 10$ per cent of the data is
rejected. Correction of the spectra for vignetting and the PSF were
peformed with the {\sc Asterix} software package. A background
spectrum was accumulated from the region inside the inner-ring of the
detector, with all sources detected above the $4 \sigma$ level masked
out and using a circular region with a radius of 90 arcsec.

The search for spectral evolution of an average QSO requires binning
the data into sufficient redshift subsamples to detect trends over the
redshift range. However, one must strike a balance between the number
of redshift bins and the resultant number of objects and photons in
each bin. Clearly the number of photons must be adequate to allow for
full spectral fitting, also, a low number of sources per bin may
result in the spectrum being dominated by one particular source,
rather then representing the spectrum of a mean source.  Details of
the redshift bins used, including the total number of background
subtracted photons in each bin, are given in table~\ref{tab:bins}. In
addition to these redshift-segregated spectra, we created a spectrum
including data from all 165 QSOs.  We note also that the photons were
distributed evenly over the fields, thus no one observation or field
will dominate the results.
 One of the two observations
of the QSF3 field, and one of the two observations
of the QSF1 field was made with the PSPC-C instrument, so with these
data we use the \verb+pspcc_gain1_256.rsp+ matrix. We use the
\verb+pspcb_gain2_256.rsp+ matrix for all other observations, as they
are made after October 1991 with the PSPC-B. Spectral fitting is
performed for spectra from each field and instrument simultaneously.

\begin{figure}
 \rotatebox{270}{\includegraphics[height=7.5cm]{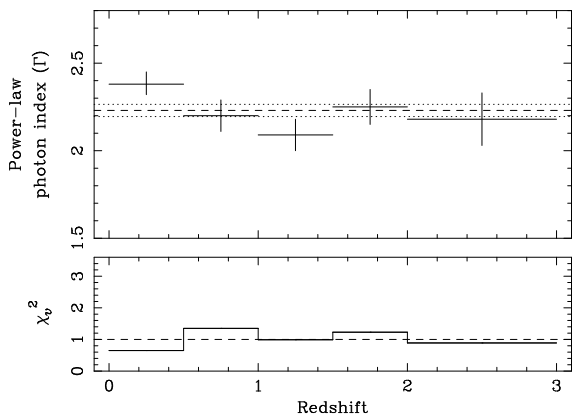}}
\caption{Results from fitting a simple power-law model over the energy
range 0.5-2.0 keV, with absorbing column fixed at the mean Galactic
value. The upper panel shows the best-fit photon indices in each
redshift bin, with the dashed line and dotted lines showing the
best-fit photon index and the $1 \sigma$ error found with stacked
data from QSOs from all redshift bins. The lower panel shows the
reduced \mchisq\ for each redshift bin.}
\label{fig:gam_po2}
\end{figure}

\begin{figure}
\rotatebox{270}{\includegraphics[height=7.5cm]{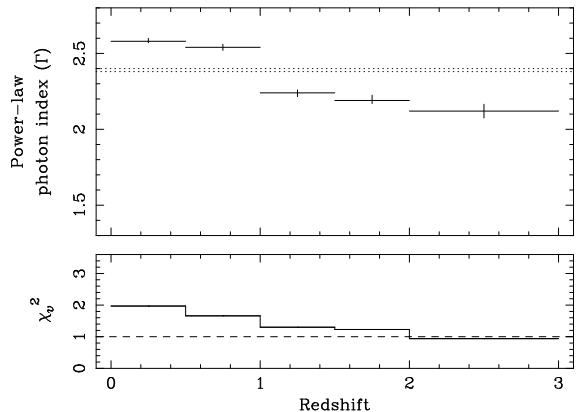}}
\caption{Results from fitting a simple power-law model over the energy
range 0.1-2.0 keV, with an absorbing column fixed at the Galactic
value. The upper panel shows the best-fit photon indices with $1
\sigma$ errors, the lower
panel the reduced \mchisq\ for each redshift bin. In the upper panel
the best-fit $\G$, from a model fit to stacked data from QSOs in all
redshift bins, is shown with a dashed line, with $1 \sigma$ errors
given by the dotted lines.}
\label{fig:gam_po}
\end{figure}

\begin{figure}
 \rotatebox{270}{\includegraphics[height=7.5cm]{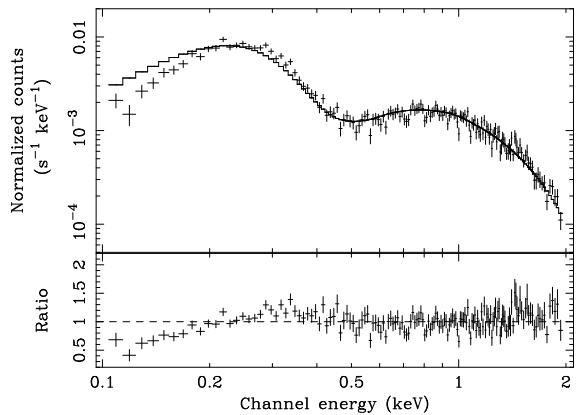}}
\caption{
Plot of the co-added spectrum from QSOs, at all redshifts,
observed with the PSPC-B instrument, with the best-fit power-law model
(see table~\ref{tab:spectra2})
shown by the solid line. The data/model ratio is given in the lower
panel, which exhibits the systematic discrepancies typical of our QSO
sample when fitted with a simple power-law model}
\label{fig:spec}
\end{figure}

\begin{figure}
 \rotatebox{270}{\includegraphics[height=7.5cm]{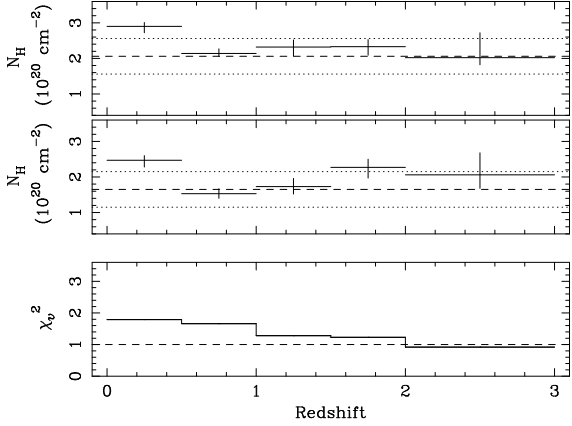}}
\caption{Free-fit absorption, found in the 0.1-2.0 keV band using a
power-law plus photoelectric absorption model, plotted against
redshift. We show the results from the PSPC-B spectra, and PSPC-C
spectra in the uppermost and middle panels respectively, with $1
\sigma$ errors.
The dashed lines are the mean Galactic values appropriate for each of
the spectra,
from Stark et al. (1992), with the dotted lines showing the
approximate uncertainty in the Stark et al. values. The lower panel
gives the reduced \mchisq\ for each model fit}
\label{fig:gam_nh}
\end{figure}

\begin{figure}
\rotatebox{270}{\includegraphics[height=7.5cm]{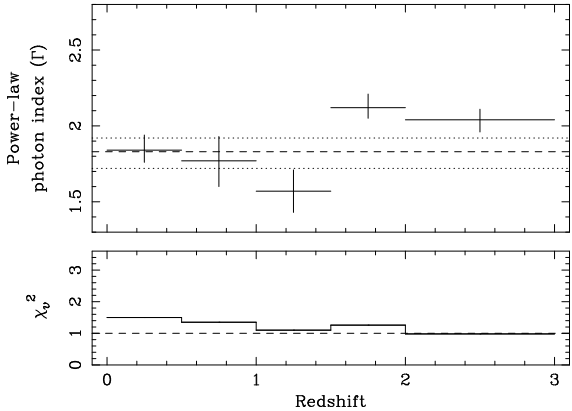}}
\caption{Best-fit power-law photon indices (along with $1 \sigma$
errors) found over the energies
$0.1-2.0$ keV, when including a redshifted black-body soft-excess 
 component in
the spectral model. The reduced \mchisq\ for each redshift bin is given
in the lower panel. In the upper panel, the dashed line denotes the
best-fit photon index found when fitting a power-law plus black-body
model to data from all the QSOs, with $1 \sigma$ errors shown by the
dotted lines.}
\label{fig:gam_pobb}
\end{figure}

\begin{figure}
\rotatebox{270}{\includegraphics[height=7.5cm]{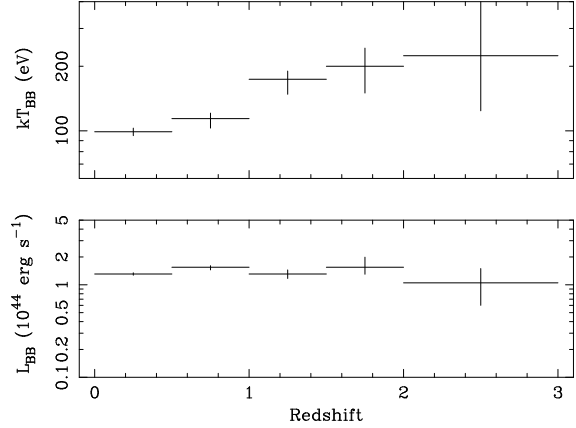}}
\caption{Best fitting black-body temperatures (top panel) and
calculated luminosities (lower panel) found for each redshift bin
whilst fixing the power-law slope to $\G = 2.0$.}
\label{fig:bbt}
\end{figure}

\section{Spectra and results}

Initially we considered the energies in the range $0.5-2.0$ keV, and
fitted a simple power-law plus absorption column model to the data,
for comparison with the results from Almaini et al. (1996). The
Galactic \mnh\ for all our fields (Stark et al. 1992, see
table~\ref{tab:fields}) is reasonably low, with a range of $\sim
1.2 - 3.0 \nunits$ and an exposure weighted 
mean of  \mnh $\approx 2.0 \x 10^{20} \nunits$. Hereinafter, unless
otherwise stated, for the Galactic \mnh\ used in model fitting, we use
appropriate $\nh$ for each of the  fields included in each
of the spectra. The
best-fit power-law indices found from the stacked photons from all
redshift bins is $\G \sim 2.23 \pm 0.07$ (table~\ref{tab:spectra2}),
with a \mchisqr\ value of 1.07 from 157 degrees of freedom, in
excellent agreement with Almaini et al. Errors quoted throughout this
paper are 90 per cent errors calculated on the basis of the
conservative assumption that all spectral parameters (including
normalisation) are interesting.

Following this we fitted the same model to data from each redshift bin
individually, resulting in best-fit photon indices presented
graphically in Fig.~\ref{fig:gam_po2}, and in tabular form in
table~\ref{tab:spectra1}. A \mchisq\ test shows that this model is
formally acceptable in each redshift bin, furthermore, we find a
non-evolving spectral slope, consistent, within errors, in each
redshift bin, with the value for $\G$ found when considering data from
all redshift bins together. These results are also consistent with the
Almaini et al. results for the QSO spectral slope as a function of
redshift [see Fig. 3b of Almaini et al. (1996)].

We then tested our data over the range $0.1-2.0$ keV with the same
model. The derived photon indices are shown in the upper panel of
Fig.~\ref{fig:gam_po}, with the \mchisqr\ given in the lower
panel. For comparison, the best-fit photon index found for data taken
from all QSOs is represented by the dashed line in the upper panel,
with the $1 \sigma$ error shown by the dotted lines.  Evolution of the
derived power-law photon index is found, in that $\G_{(z<1.0)} > 2.50$
and $\G_{(z>1.0)} \le 2.25$. To establish the significance of this
difference in spectral slope, we performed spectral fits on two
further spectra, including data from QSOs in the redshift range
$0.0-1.0$ and $1.0-3.2$ respectively (for results see
table~\ref{tab:spectra2}). This difference is significant at over the
$8.5 \sigma$ level. It is, however, easily seen that this model
provides an increasingly inadequate description of the data for
smaller redshifts, indeed a \mchisq\ test rejects the model fit in the
two lower redshift bins to at least the $99.9$ per cent level.

The typical shape of a spectrum that rejects this model is shown in
Fig.~\ref{fig:spec}, the residuals (lower panel) clearly indicating
a systematic difference between the data and the model at energies
$\simlt0.5$ keV. To determine whether the data requires absorption in
excess of the Galactic value, as seen in a small number of AGN
(Ciliegi \& Maccacaro 1996), we allowed the absorbing column density
to be a free parameter in the fit, again performing the fit for both PSPC-B
and PSPC-C spectra simultaneously,  and with independent column
densities. Fig.~\ref{fig:gam_nh} shows how the best fitting column
density varies with redshift for the PSPC-B spectra (uppermost panel)
and the PSPC-C spectra (middle panel), with the dashed line
representing the appropriate weighted mean Galactic \mnh\, and the
dotted lines the estimated uncertainty in the mean. The total
absorbing column is consistent with the galactic value in all the
redshift bins. Neither  inclusion of \mnh\ as a free parameter nor
including an intrinsic warm-absorber  
reduce the \mchisq\ significantly, as determined by an F-test
(Bevington \& Robinson 1994), in any of the redshift bins. 

In an attempt to fit the $0.1-2.0$ keV spectra, and guided by results
from nearby AGN which have a  spectral upturn towards low energies
 (eg  Turner and Pounds, 1989, Fiore et al. 1994), 
 we tried a two-component model consisting of a
power-law plus a redshifted black-body, the derived photon indices and
\mchisqr\ are shown in Fig.~\ref{fig:gam_pobb} and
table~\ref{tab:spectra1}. We see a significant reduction in the
\mchisq\ due to the inclusion of this soft-excess even though the
model is still formally rejected in the lowest redshift bin. However this
result is probably not unreasonable, as we have co-added the photons
from many sources at slightly different redshifts (and having different
physical parameters and geometries), thus a single black-body model
may not provide a perfect description of the data.  We find that the
power-law slope derived when including a black-body soft-excess is, in
each redshift bin, consistent within the 90 per cent errors with $\G
\approx 1.8 \pm 0.2$, the value found for the entire sample of QSOs
together. This value is consistent, within the 90 per cent error bars,
with the intrinsic AGN power-law index of $\G \sim 1.9$ (Nandra \&\
Pounds 1994), suggesting that the hard power-law continues down to
below $\sim 0.5$ keV in the QSO rest frame.  The best-fit temperatures of the
black-body components (kT $\sim 100 - 200$ eV) are in general
agreement with those found for quasars (Rachen et al. 1996) and nearby
radio-quiet QSOs (Fiore et al. 1994).  Modelling the soft-excess as a
thermal bremsstrahlung, a hot plasma, or an additional power-law 
  yields similar \mchisq\ results
over the five redshift bins, and derived power-law indices that,
again, are around $\G \sim 1.8$.

\begin{table*}
\begin{center}
\begin{tabular}{ccccccc}
Redshift& \multicolumn{2}{c}{Model: PL} 
        & \multicolumn{2}{c}{Model: PL}
        & \multicolumn{2}{c}{Model: PL$+$BB} \\
range   & \multicolumn{2}{c}{$0.5-2.0$ keV}
        & \multicolumn{2}{c}{$0.1-2.0$ keV}
        & \multicolumn{2}{c}{$0.1-2.0$ keV}\\
        & $\G $ & \mchisqr (\mnu) & $\G $ 
        & \mchisqr (\mnu) & $\G $ & \mchisqr (\mnu) \\ \hline
0.1-3.2 &$ 2.23^{+0.07}_{-0.07} $&$ 1.07 (157) $
        &$ 2.39^{+0.02}_{-0.02} $&$ 2.12 (231) $
        &$ 1.83^{+0.17}_{-0.19} $&$ 1.74 (228) $ \\
0.1-1.0 &$ 2.32^{+0.10}_{-0.11} $&$ 1.25 (97)  $
        &$ 2.60^{+0.02}_{-0.03} $&$ 2.68 (167) $
        &$ 1.80^{+0.21}_{-0.23} $&$ 2.09 (164) $ \\
1.0-3.2 &$ 2.15^{+0.11}_{-0.10} $&$ 1.20 (108) $
        &$ 2.24^{+0.03}_{-0.04} $&$ 1.60 (168) $
        &$ 1.90^{+0.28}_{-0.15} $&$ 1.54 (165) $ \\
\end{tabular}
\caption{Details of spectral fitting results. Errors quoted are at the
90 per cent confidence level.
}
\label{tab:spectra2}
\end{center}
\end{table*}

\begin{table*}
\begin{center}
\begin{tabular}{cccccccc}
  & Redshift
  & \multicolumn{2}{c}{Model: PL}
  & \multicolumn{2}{c}{Model: PL}
  & \multicolumn{2}{c}{Model: PL$+$BB} \\
  & range
  & \multicolumn{2}{c}{$0.5-2.0$ keV}
  & \multicolumn{2}{c}{$0.1-2.0$ keV}
  & \multicolumn{2}{c}{$0.1-2.0$ keV} \\
  & & $\G $ & \mchisqr (\mnu) & $\G $
  & \mchisqr (\mnu) & $\G $ & \mchisqr (\mnu) \\ \hline
1 & $0.0 - 0.5$
  & $ 2.38^{+0.13}_{-0.12} $ & $ 0.65 (58)  $ 
  & $ 2.53^{+0.03}_{-0.03} $ & $ 2.23 (224) $
  & $ 1.84^{+0.16}_{-0.19} $ & $ 1.34 (217) $ \\
2 & $0.5-1.0$
  & $ 2.20^{+0.17}_{-0.17} $ & $ 1.35 (47)  $
  & $ 2.54^{+0.05}_{-0.05} $ & $ 1.42 (159)  $
  & $ 1.77^{+0.32}_{-0.33} $ & $ 1.12 (149)  $ \\
3 & $1.0-1.5$
  & $ 2.09^{+0.18}_{-0.18} $ & $ 0.99 (60)  $
  & $ 2.21^{+0.04}_{-0.05} $ & $ 1.31 (180) $
  & $ 1.57^{+0.27}_{-0.27} $ & $ 1.05 (169)  $ \\
4 & $1.5-2.0$
  & $ 2.25^{+0.20}_{-0.20} $ & $ 1.23 (45)  $
  & $ 2.14^{+0.07}_{-0.06} $ & $ 0.99 (128)  $
  & $ 2.11^{+0.18}_{-0.14} $ & $ 0.95 (117)  $ \\
5 & $2.0-3.2$
  & $ 2.18^{+0.30}_{-0.30} $ & $ 0.89 (19)  $
  & $ 2.08^{+0.10}_{-0.10} $ & $ 1.44 (47)  $
  & $ 2.04^{+0.14}_{-0.15} $ & $ 1.09 (38)  $ \\
\end{tabular}
\caption{Details of spectral fitting results in five redshift
bins. Errors quoted are at the 90 per cent confidence level.
}
\label{tab:spectra1}
\end{center}
\end{table*}

While we note that our description of the soft-excesses found in the
binned spectra as a simple single-temperature black-body is simply
an expedient to obtain a better statistical representation of the
data and as such may not be fully representative of the true spectra
of individual objects, or the underlying physical properties, it is
of interest to examine the behaviour of the black-body parameters
as a function of redshift bin.
In Fig.~\ref{fig:bbt} and table~\ref{tab:bb} we present the best fit
black-body temperature and the luminosity of the black-body component
as a function of redshift (now fixing the power-law photon index at 2
in order to better constrain the black-body properties).  It is clear
that the black-body temperature, while consistent with measurements
for nearby objects of similar luminosity (Fiore et al. 1994) at zero
redshift, increases in temperature to higher redshifts, rising from
$\sim 100$ eV in the most local bin to of order 200 eV at redshifts
$\sim 2$.  In contrast, the mean luminosity of the black-body per
object remains approximately constant.
 
\begin{table}
\begin{center}
\begin{tabular}{cccc}
  &  Redshift & $kT_{BB}$       & $L_{BB} $ \\ 
  &  range    &   (eV)              & $(10^{44} \lunits)$ \\ \hline
1 & $0.0-0.5$ & $101^{+5}_{-5}$      & $1.31^{+0.08}_{-0.08}$ \\
2 & $0.5-1.0$ & $119^{+14}_{-15}$   & $1.55^{+0.12}_{-0.19}$ \\
3 & $1.0-1.5$ & $174^{+25}_{-32}$   & $1.31^{+0.27}_{-0.27}$ \\
4 & $1.5-2.0$ & $194^{+79}_{-91}$  & $1.55^{+0.87}_{-0.50}$ \\
5 & $2.0-3.2$ & $274^{+80}_{-123}$ & $1.05^{+0.90}_{-0.90}$ \\
\end{tabular}
\caption{Temperature and luminosity of the best-fit black-body
component. Errors are 90 per cent. The power-law component was fixed
at $\G = 2$ to allow better constraint of the black-body parameters.
}
\label{tab:bb}
\end{center}
\end{table}

We first ask whether the effect could be some instrumental artefact, as
the black-body temperature in the observer's frame remains roughly
constant. A systematic error in the calibration of the effective area,
for example, could cause such an effect.  We can rule out this
possibility  by noting that the fractional contribution from the
soft excess relative to the power-law component varies with redshift
from bin to bin, while a systematic uncertainty or error in the
calibration would give a constant fractional contribution.
Another possibility is that the results we obtain are an
artefact of our assumption that the spectra of QSOs
within a redshift bin can be represented by  a single value of 
$\Gamma$. For example coadding a step and a flat spctrum 
 would result in a steep and flat spectrum at low energies 
 and high energies respectively i.e. in a concave spectrum. 
 In principle, 
 deriving the spectra of individual QSOs would circumvent this problem. 
 Fiore et al. (1994) find that a large fraction of 
 the QSOs in their sample (four out of six) reveal 
 a soft excess at low PSPC energies.  In contrast Laor et al. (1997) 
 find that the PSPC spectra of the majority of their  LBQS QSOs are 
consistent with
 a single power-law to within 30\%. Unfortunately, 
 our objects typically are faint 
 and we cannot derive individual spectra. 
 However, as an exercise we have derived 
 individually the spectra  of the two brightest QSOs in the QSF3 field.
 The column density was fixed at the Galactic value $N_H=1.7\times 10^{20}
 \rm cm^{-2}$.  
 For RXJ0342.6-4404 (z=0.377) we obtain $\Gamma=3.18^{+0.08}_{-0.08}$  
  for a single power-law fit 
 ($\chi^2= 59.9/51$ dof). 
 The addition of a black-body component (0.06 keV) is 
 significant at the 99 per cent confidence level ($\chi^2= 51.0/49$ dof)
 with $\Gamma=2.58^{+0.25}_{-0.21}$. 
 In contrast RXJ0342.0-4403 (z=0.635) can be fit with a single power-law 
 ($\Gamma =2.9^{+0.20}_{-0.10}$) without the need of an additional component
 ($\chi^2=40.3/38$ dof). 
 It is evident that a conclusive answer 
 on the origin of the soft excess cannot be given 
 with the photon statistics curently available. 
 Finally, it is possible that a range  of  
 individual QSO spectra  could result in
artificially induced behaviour which varies with redshift due to 
sampling different energy ranges in each redhsift bin. 
If for example QSOs had a range of spectral
indices then sampling at different energy ranges would bias the
results in that using  higher energy bands flatter spectra would be
preferentially be sampled giving rise to an apparent flattening of
spectrum with redshift. Such an explanation for the behaviour observed
here seems contrived and unlikely given the lack of variation in
spectra
in bins with redshifts greater than $\sim 1$ .

Next, we consider the possible physical origin of the  
 spectral upturn. The soft
excesses observed in AGN spectra are often ascribed to the high energy
end of the big blue bump seen in the optical/UV region of the spectrum
and thought to originate in the accretion disk (Walter \& Fink 1993,
Turner \& Pounds 1989). The properties of the accretion disk are then
dominated by the mass accretion rate and mass of the central
black-hole. In the simplest models of the emission from the accretion
disk, the spectrum is a multi-temperature black-body with the maximum
temperature component coming from the inner disk radii.  
The behaviour of the soft-excess parameters presented here certainly
suggest that the emission is more complex than would be expected in a
thin accretion disk. The high temperatures found do not 
 fit well with the above models. Moreover, the 
 constant luminosity of the soft excess seen
in the {\it ROSAT} band while the power law component increases by a
factor of 20-30 is more consistent with up-scattering than with
thermal emission from a disk powered by potential energy loss. We note
that the constant luminosity of our measured soft excesses is similar
to that found by Saxton et al. (1993) for a sample of nearby
QSOs. They found that the soft excess luminosity saturated above a
power law luminosity of $\sim 10^{44} \lunits$, the mean
luminosity in our lowest redshift bin. They postulated that this was
due to a {\it decreasing} temperature of the soft-excess. Such an
explanation is contradicted by our results, which are more in keeping
 with an increasing mass-accretion rate rather than black-hole mass
being the important factor.

Regardless of the origin of the spectral upturn at low 
 energies, the average QSO spectrum  
 has important implications for the origin of the 
 XRB. Boldt et al. (1987) argue that as the AGN have 
 hard (2-10 keV) X-ray spectra,  
  significantly steeper than the spectrum of the XRB 
 in the same band (Marshall et al. 1980), 
 they cannot produce the bulk of the XRB. 
 This spectral paradox appears to extend at the lower ROSAT 
 energies. The XRB in the 0.5-2 keV band also has this  flat spectral index, 
 $\Gamma\sim 1.5$, (Georgantopoulos et al. 1996, Vecchi et al. 1999).  
 All previous studies of the QSO  spectra 
 in the {\it ROSAT} band   
 (eg Schartel et al. 1996, Laor et al. 1997) obtain 
 spectral indices much steeper than the XRB spectrum 
 in the same band. However, these contain
 mainly bright, low redshift QSOs.    
 In contrast, our sample contains  {\it typical} QSOs 
 in the sense that our objects  constitute 
 a large fraction (over $\sim 50$ per cent at 1 keV) of the XRB
 and they cover a wide range of redshift and luminosity. 
 Our results clearly show that the soft X-ray selected QSOs 
 cannot produce the bulk of the soft XRB. 
 Another population with a flat X-ray spectrum is needed 
  at faint fluxes. This population may be associated with 
 AGN obscured in X-ray wavelenghts; 
 some examples of this population at low redshift 
 ($z<0.5$)  may have been already identified 
 in deep {\it ROSAT} pointings (Schmidt et al. 1998).

\section{Conclusions}

We derived the average QSO spectrum and its evolution with redshift,
using a sample of 165 QSOs from 7 deep {\it ROSAT} fields ($f_{\rm x}
>4 \x 10^{-14} \funits$). We find strong evidence ($>8 \sigma$) for
spectral evolution, in that the $0.1-2.0$ keV spectra flatten from $\G
\sim 2.6$ at low redshifts, to $\G \sim 2.1$ at $z \sim 2.4$.  At high
redshifts a power-law model with Galactic absorption describes the
stacked spectra well, with the power-law index in rough agreement with
the slope found from the $0.5-2.0$ keV spectra.  However, for z$<1.5$,
the spectra of the QSOs in the $0.1-2.0$ keV band cannot be fitted by
a single power-law, with Galactic absorption, in contrast with some 
 other {\it ROSAT} results on low redshift QSOs (eg Laor et al. 1997) 
 Furthermore, we find that these deviations from
 a simple power-law cannot be modelled by excess \mnh\ absoption along
the line of sight.  The above suggest the presence of a soft excess
below 0.5 keV in the QSO rest-frame. Indeed, when we add a black-body
component to model the soft excess, we obtain a significant reduction
in the \mchisq\ for the low redshift bins. However this model is still
rejected in the lowest redshift bin, probably because a single
temperature black-body cannot represent adequately the coadded soft
excess of all QSOs in one bin. The resulting power-law component has a
photon index $\G \sim 1.8 \pm 0.2$, and shows no evidence of evolution
with redshift. This spectral index is not inconsistent, within the 90
per cent errors, with both the canonical AGN spectral index of $\G
\sim 1.7$ observed in the $2-10$ keV band, and the intrinsic spectral
slope of $\Gamma \sim 1.9$, suggesting that the hard X-ray power-law may
continue well into the {\it ROSAT} band, down to below $\sim 0.5$
keV. The soft excess probably evolves with redshift, maintaining a
constant luminosity but increasing in temperature from 100 eV locally,
to 200 eV at $z \sim 2$.
We caution that the black-body model 
is simply an expedient to obtain a better statistical 
 representation of  the data. It is  possible that 
 the soft excess observed may be just an artefact of the  
 coaddition of spectra with a large range of spectral indices. 
 Regardless of the origin of the 
 soft excess, the average QSO spectrum has important 
 implications for the
 origin of the XRB. It is shown here that the average spectrum of the 
 faint, high redshift QSOs is much steeper than the spectrum of the 
 XRB in the soft 0.1-2 keV band, extending the 
 spectral paradox to low energies. 
 Future missions such as XMM are expected 
 to bring a breakthrough in the spectral studies 
 of high redshift, faint QSOs due 
 to the broad energy coverage 
 and high effective area.

\section{acknowledgments}
AJB acknowledges the receipt of a University of Leicester
studentship. IG and OA acknowledge the support of PPARC.

\end{document}